\documentclass{ws-procs961x669}
\usepackage{ws-toc,ws-multind}
\usepackage{enumitem}

\makeindex{author}

\begin{document}

\title{Impact of spectator fields and non-minimal couplings in spontaneous baryogenesis}

\author{Mattia Dubbini}
\index{author}{Dubbini, M.}

\address{Universit\`a di Camerino, Via Madonna delle Carceri, Camerino, 62032, Italy.\\E-mail: mattia.dubbini@unicam.it}

\begin{abstract}

We investigate the model of spontaneous baryogenesis, considering two extensions to the background paradigm. Firstly, we introduce a non-minimal coupling between gravity and the inflaton, increasing the effective mass squared of the latter. In this way, the inflaton decays more likely into fermion-antifermion pairs during reheating, through baryon-number violating processes. Accordingly, we obtain an overall baryon asymmetry consistent with cosmological observations. Then, we consider a complex scalar spectator field interacting with the inflaton through a biquadratic coupling and non-minimally with gravity, and analyze the impact in terms of baryon asymmetry production. In this scenario too, the background model results significantly enhanced, but the predicted baryon-to-entropy ratio remains smaller than the experimental data.

\end{abstract}

\section{Introduction}

The baryon-antibaryon asymmetry of the universe is an open problem in modern cosmology. Indeed, cosmological observations suggest that the universe is prevalently made by baryons, while the antimatter component is much less present. The observed asymmetry, however, cannot be explained by the Standard Cosmological Model.

The theories that explain how baryons and antibaryons were created in the early universe are called theories of baryogenesis \cite{Luongo:2021gho, Dubbini:2025xvg}. Since 1967, when A.D. Sakharov established the three criteria for yielding baryon asymmetry, many models have been formulated. Among them, we can find electroweak baryogenesis \cite{Farrar:1993hn}, baryogenesis via leptogenesis \cite{Fong:2012buy}, and almost purely geometric scenarios, such as gravitational baryogenesis \cite{Arbuzova:2023rri}. However, one of the most compelling model is spontaneous baryogenesis \cite{Dolgov:1994zq, Dolgov:1996qq}, which assumes that the asymmetry is produced from the spontaneous breaking of the $U(1)_B$ global symmetry and the consequent baryon-number violating decays of the inflaton during reheating.

In this work, we study two different extensions to the background model of spontaneous baryogenesis. Firstly, we generalize the minimal model to a non-minimal one, introducing a non-minimal coupling term between gravity and the inflaton \cite{Dubbini:2025jjz}. Then, we consider a complex scalar spectator field interacting with the inflaton through a biquadratic coupling and non-minimally with gravity \cite{Dubbini:2025hcw}. We show that both the scenarios predict significant enhancements in terms of baryon asymmetry production, even though only the non-minimal framework can reproduce the experimental data.

\section{The background model of spontaneous baryogenesis}

In the framework of spontaneous baryogenesis:

\begin{itemize}[leftmargin=*]

\item[-] baryon number is conserved above a characteristic symmetry-breaking scale $f$;

\item[-] below $f$, the global $U(1)_B$ symmetry is spontaneously broken, and the related pseudo Nambu-Goldstone boson is identified with the inflaton;

\item[-] the inflaton decays into fermions Q and L -- respectively carrying baryon number and not -- and their antiparticles via baryon-number and CP violating interactions, yielding a net baryon asymmetry in agreement with Sakharov criteria.

\end{itemize}

The corresponding theoretical setup is described by the Lagrangian density
\begin{equation}
\begin{split}
\mathcal{L}_0&=\frac{f^2}{2}(\partial_{\mu}\theta)(\partial^{\mu}\theta)+\overline{Q}(i\gamma^{\mu}\partial_{\mu}-m_Q)Q+\overline{L}(i\gamma^{\mu}\partial_{\mu}-m_L)L+\\&+\frac{gf}{\sqrt{2}}(\overline{Q}L+\overline{L}Q)-(\partial_{\mu}\theta)J^{\mu}-V(\theta),
\end{split}
\label{brokenphaseLagrangian}
\end{equation}
where $V(\theta)$ is a generic inflationary potential and $J^{\mu}=\overline{Q}\gamma^{\mu}Q$ is the baryonic current. Solving the inflationary dynamics in a homogeneous and isotropic background -- neglecting the cosmic expansion -- yields $\theta_0(t)=\theta_Ie^{-\frac{\Gamma t}{2}}\cos(\Omega t)$, where $\Omega$ is the renormalized mass of the inflaton and $\Gamma=g^2\Omega/(8\pi)$ is its global decay rate ($g\ll 1$).

According to Eq. (\ref{brokenphaseLagrangian}), the inflaton can decay via $\theta\to Q+\overline{L}$ and $\theta \to \overline{Q}+L$, producing baryons and antibaryons respectively. As the responsible interaction violates the CP symmetry, the two channels have different decay rates, yielding a non-null number density of baryons minus antibaryons $n_B=n(Q,\overline{L})-n(\overline{Q},L)$. Computing the decay amplitudes for each of the two channels, the overall baryon-to-entropy ratio turns out to be
\begin{equation}
\eta_0\equiv\frac{n_B^{(0)}}{s}=3\cdot10^{-3}\frac{g^3}{g_*^{\frac{1}{4}}}\biggl(\frac{M_{\text{Pl}}}{f}\biggl)^{\frac{3}{2}}\sqrt{\frac{f}{\Omega}}\biggl(\frac{1-\epsilon^2}{1+\epsilon^2}\biggl)^2,
\label{baryon-to-entropyBK}
\end{equation}
even considering mass-mixing between fermions and cosmic expansion.

\section{Non-minimal coupling between inflaton and gravity}

We extend the Lagrangian in Eq. (\ref{brokenphaseLagrangian}) to the following theoretical model
\begin{equation}
\mathcal{S}=\int d^4x\sqrt{-g}\left[-\frac{M^2_{\text{Pl}}}{16\pi}R-\frac{1}{2}f^2\xi\theta^2R+\mathcal{L}_0\right]
\label{actionnon-minimal}.
\end{equation}
The non-minimal coupling term modifies the effective mass squared of the inflaton according to $\Omega^2\to \Omega^2+\xi R$, affecting the decay amplitudes and consequently the baryon asymmetry. In particular, evaluating the Ricci scalar, at leading order we find $R=\Upsilon \dot{\theta}$, with $\Upsilon\sim f$. 

We solve the inflationary dynamics using perturbation method up to first order in $\xi\ll1$, taking $\theta=\theta_0+\xi\theta_1$. At leading order, on average $\dot{\theta}=\dot{\theta_0}<0$, and thus $R<0$. Accordingly, we expect a positive correction to the inflaton mass squared for $\xi<0$ and negative for $\xi>0$, enhancing baryogenesis in the first case while suppressing it otherwise. 

\begin{figure*}
\centering
{\hfill
\includegraphics[width=0.515\hsize,clip]{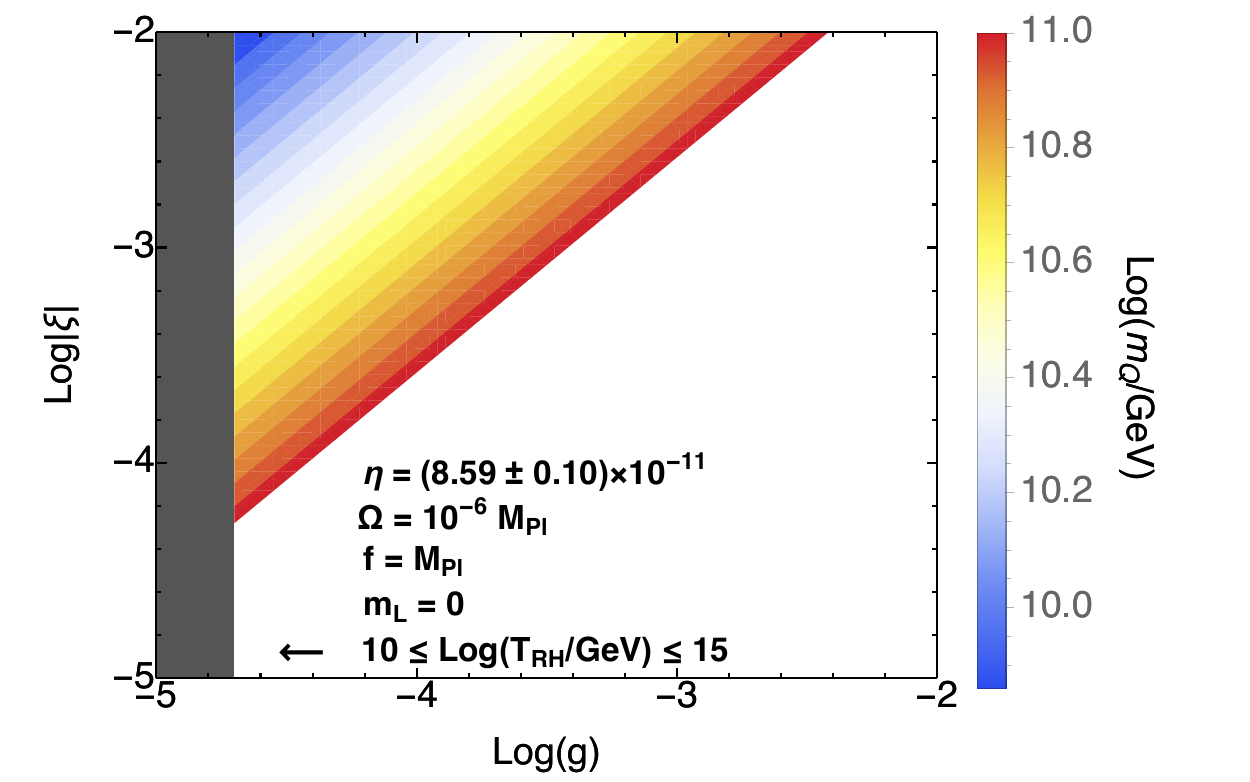}
\hfill
\includegraphics[width=0.475\hsize,clip]{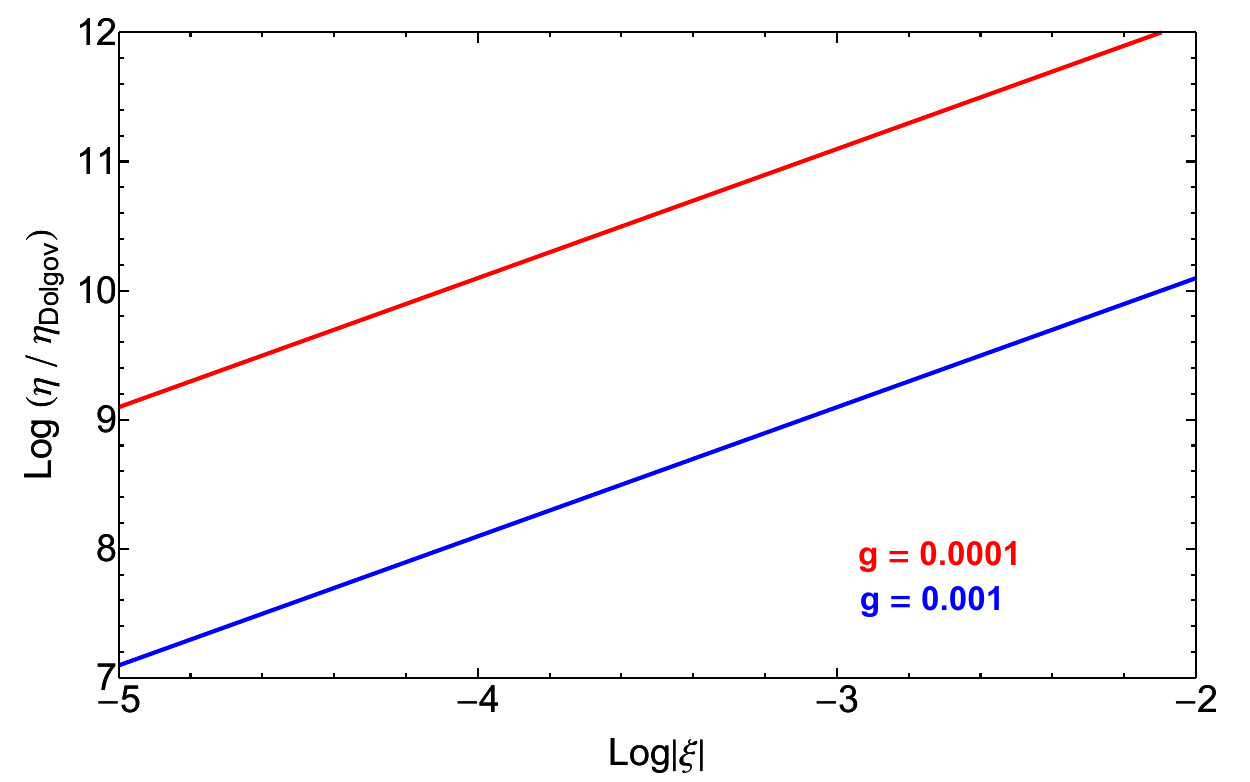}
\hfill}
\caption{\emph{Left}: contour plot of the allowed region of the $\log{(g)}-\log{|\xi|}$ plane, compatible with the observational constraint $\eta=(8.59\pm0.10)\times 10^{-11}$. \emph{Right}: ratio between our $\eta$ from Eq.~\eqref{baryon-to-entropynon-minimal} and the background's result as function of $\xi$, for $g=0.0001$ and $g=0.001$.}
\label{fig1}
\end{figure*}

Finally, considering the mass-mixing and the cosmic expansion, we obtain the following baryon-to-entropy ratio
\begin{equation}
\eta=\eta_0\bigg{(}1-\frac{16\pi^2\Xi}{3g^4}\bigg{)},\quad \text{with}\quad \Xi=\frac{\xi\theta_I\Upsilon}{\Omega}.
\label{baryon-to-entropynon-minimal}
\end{equation}
We show that the smaller $g$, the greater the improvement in the baryon asymmetry due to the non-minimal coupling (right panel of Fig. \ref{fig1}). Moreover, our result turns out to be able to reproduce the experimental value of the baryon-to-entropy ratio for parameters compatible with the perturbative regime (left panel of Fig. \ref{fig1}).

\section{Complex scalar spectator field}

As a second improvement of the background paradigm, we extend the Lagrangian model in Eq. (\ref{brokenphaseLagrangian}) considering the following action
\begin{equation}
\mathcal{S}=\int d^4x\sqrt{-g}\bigg{[}-\frac{M^2_{\text{Pl}}}{16\pi}R+\mathcal{L}_0+\mathcal{L}_{\phi}\bigg{]}
\label{actionspectator},
\end{equation}
where $\mathcal{L}_{\phi}$ is a Lagrangian density describing a complex scalar field interacting with the inflaton through a biquadratic coupling and non-minimally with gravity,
\begin{equation}
\mathcal{L}_{\phi}=(\nabla^{\alpha}\phi^*)(\nabla_{\alpha}\phi)-m^2_{\phi}\phi^*\phi-\xi f^2\theta^2\phi^*\phi-\sigma R\phi^*\phi.
\label{spectatorLagrangian}
\end{equation}
As the real and imaginary parts of $\phi$ satisfy the same equation of motion, we name $\phi\equiv\phi_R=\phi_I$. We obtain a system of coupled second order differential equations for $\phi$ and $\theta$, that we solve using perturbation theory and neglecting initially the expansion of the universe.

\begin{itemize}[leftmargin=*]

\item[-] We consider $\phi=\phi_{0\xi}$, neglecting terms of first or higher order in $\xi$, as in the inflaton equation these would be an $\mathcal{O}(\xi^3)$. Differently, we take $\theta=\theta_0+\xi\theta_{1\xi}$.

\item[-] The resulting equation for $\phi_{0\xi}$ is at order zero in $\xi$ and up to first order in $\sigma$. Therefore, we consider $\phi_{0\xi}=\phi_{0\xi,0\sigma}+\sigma\phi_{0\xi,1\sigma}$ and solve up to first order in $\sigma$.

\item[-] We put the full solution $\phi_{0\xi}$ in the inflaton equation and solve the inflationary dynamics up to first order in $\xi$.

\end{itemize}

\begin{figure*}
\centering
\includegraphics[width=0.70\textwidth,clip]{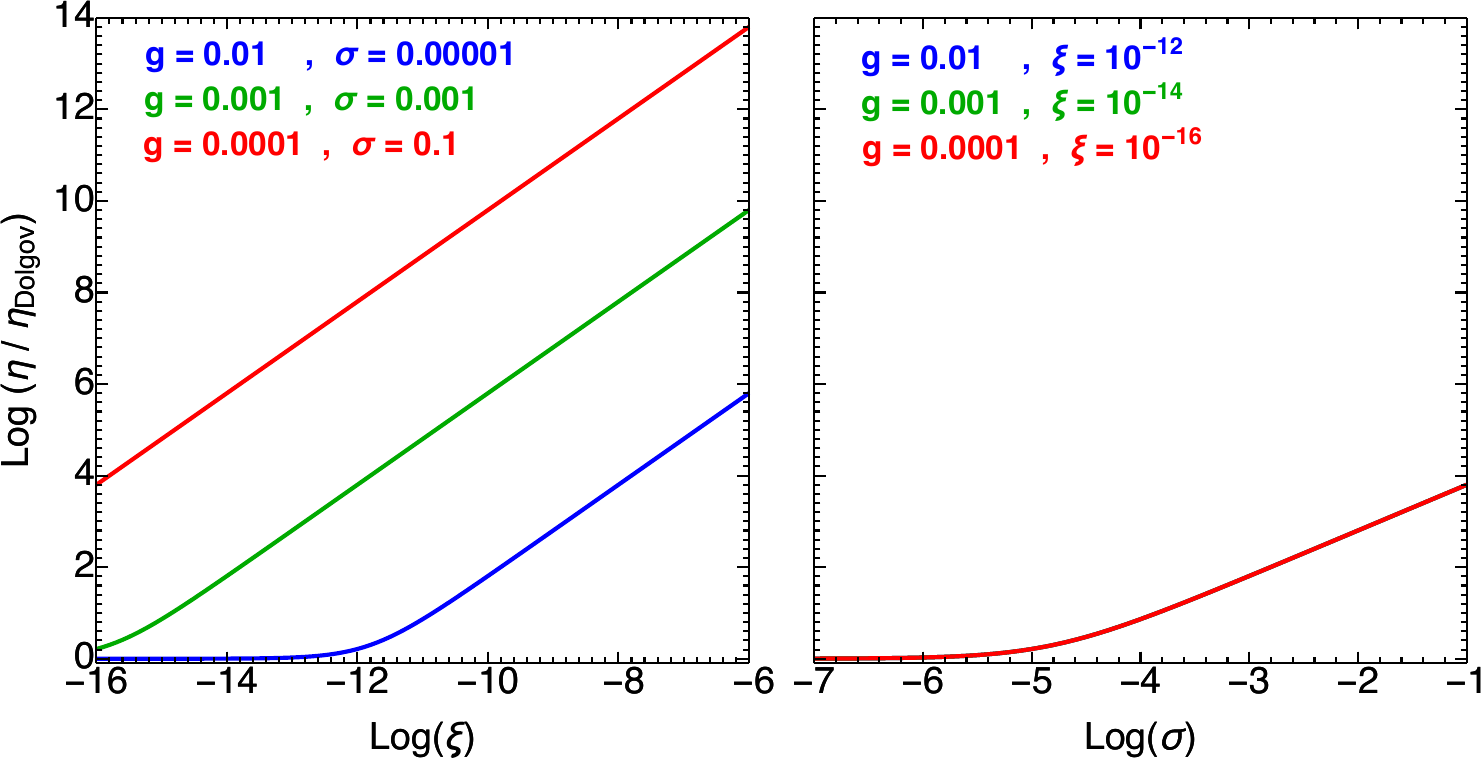}
\vspace{0.4cm}
\includegraphics[width=0.80\textwidth,clip]{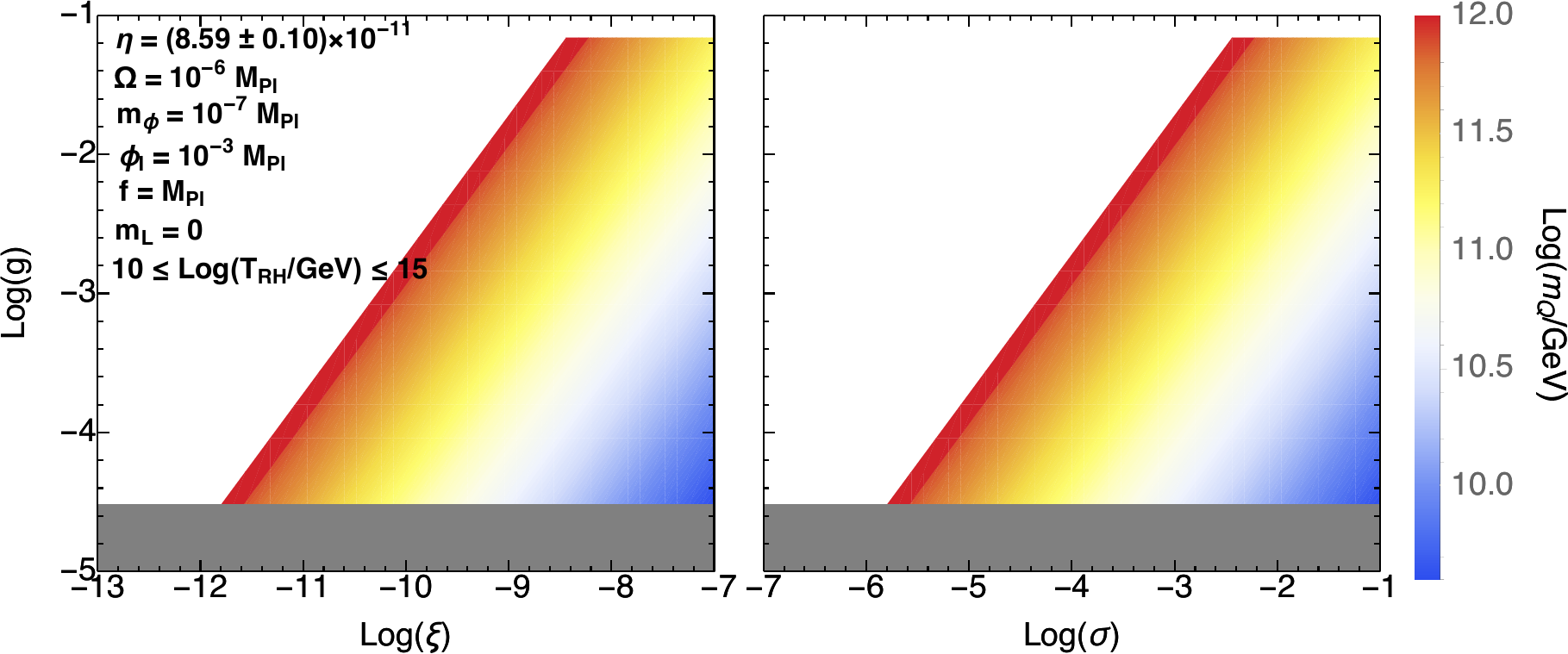}
\caption{\emph{Top:} ratio between $\eta$ from Eq.~\eqref{baryon-to-entropyspectator} and the background's result as function of $\xi$ and $\sigma$, for fixed values $g=0.01$, $g=0.001$, $g=0.0001$, and $\phi_I=10^{-3}f$. \emph{Bottom:} contour plot of the allowed region in the $\log(g)-\log(\xi)$ and $\log(g)-\log(\sigma)$ planes, compatible with the observational constraint $\eta=(8.59\pm0.10)\times10^{-11}$; gray areas are excluded by the lower bound $T_{\rm RH}=10^{10}$~GeV.
}
\label{fig2}
\end{figure*}

The first order correction ($\theta_{1\xi}$) to the background solution ($\theta_0$) provides the following baryon-to-entropy ratio
\begin{equation}
\eta=\eta_0\bigg{[}1+\Xi\bigg{(}1+\frac{64\pi^2\alpha\Sigma}{3g^4}\bigg{)}\bigg{]},\quad \text{with}\quad \Xi=\frac{\xi\phi_I^2}{\Omega^2}\quad \text{and}\quad \Sigma=\frac{\sigma\Upsilon\theta_I}{m_{\phi}}.
\label{baryon-to-entropyspectator}
\end{equation}
For parameters compatible with perturbative regime, we obtain significant improvements in terms of baryon asymmetry production with respect to the background model (top panel of Fig. \ref{fig2}). In particular, we find two first-order corrections to $\eta_0$, among which the dominant ($\propto \xi\sigma$) is due to the non-minimal coupling between gravity and the inflaton. Nevertheless, Eq. (\ref{baryon-to-entropyspectator}) is still not sufficient to reproduce the experimental value of the baryon-to-entropy ratio (bottom panel of Fig. \ref{fig2}), suggesting the need of further improvements.

\section{Conclusions and perspectives}

In this work, we extended the model of spontaneous baryogenesis considering two different scenarios. Firstly, we introduced a non-minimal coupling term between gravity and the inflaton. We showed that this term significantly enhances baryogenesis, and accordingly the model can reproduce the experimental value of the baryon asymmetry. Then, we considered a complex scalar spectator field non-minimally coupled to gravity and through a biquadratic interaction to the inflaton. Also in this case, we obtained a remarkable improvement in terms of baryon asymmetry, but not sufficient to reproduce cosmological observations.

Future works comprehend further extensions of the spontaneous baryogenesis framework, such as including the electromagnetic sector through a coupling with the inflaton. Another intriguing possibility is the introduction of a disformal non-minimal coupling between inflaton and gravity, as a generalization of the conformal one studied in this work. It would also be worthwhile to replace the complex scalar spectator field with a fermion, to evaluate its impact on baryogenesis. Lastly, a more realistic treatment of spontaneous baryogenesis would incorporate quantum chromodynamics, possibly unveiling new mechanisms of asymmetry generation.

\bibliography{bibliography}

\end{document}